\documentclass[twocolumn,pre,showpacs,eqsecnum]{revtex4}

\usepackage{dcolumn}
\usepackage{amsfonts}
\usepackage{amsmath}
\usepackage{amssymb}
\usepackage{bm}

\newif\ifpdf
\ifx\pdfoutput\undefined
\pdffalse 
\else
\pdfoutput=1 
\pdftrue \fi

\ifpdf
\usepackage[pdftex]{graphicx}
\else
\usepackage{graphicx}
\fi

\begin{document}

\ifpdf \DeclareGraphicsExtensions{.jpg,.pdf,.tif} \else
\DeclareGraphicsExtensions{.eps,.jpg} \fi

\setlength{\arraycolsep}{0.7mm}

\newcommand{\brm}[1]{\bm{{\rm #1}}}
\newcommand{\tens}[1]{\underline{\underline{#1}}}

\title{Dynamics of nematic elastomers}

\author{Olaf Stenull}
\affiliation{ Department of Physics and Astronomy,
University of Pennsylvania, Philadelphia, Pennsylvania
19104, USA }

\author{T.~C.~Lubensky}
\affiliation{ Department of Physics and Astronomy,
University of Pennsylvania, Philadelphia, Pennsylvania
19104, USA }

\vspace{10mm}
\date{\today}

\begin{abstract}
\noindent We study the low-frequency, long-wavelength
dynamics of soft and semi-soft nematic elastomers using two
different but related dynamic theories. Our first
formulation describes the pure hydrodynamic behavior of
nematic elastomers in which the nematic director has
relaxed to its equilibrium value in the presence of strain.
We find that the sound-modes structure for soft elastomers
is identical to that of columnar liquid crystals. Our
second formulation generalizes the derivation of the
equations of nematohydrodynamics by the Harvard group to
nematic elastomers.  It treats the director explicitly and
describes slow modes beyond the hydrodynamic limit.
\end{abstract}


\maketitle

\section{Introduction}
\noindent Nematic elastomers
(NEs)~\cite{WarnerTer2003,FinKoc81,deGennes,Terentjev99}
are rubbery materials with the macroscopic symmetry
properties of nematic liquid
crystals~\cite{deGennesProst93,Chandrasekhar92}. In
addition to the elastic degrees of freedom of ordinary
rubber, nematic elastomers possess the internal,
orientational degree of freedom of liquid crystals. Since
NEs are amorphous solids rather than fluids, their
mechanical properties differ significantly from those of
standard nematic liquid crystals. The interplay between
elastic and orientational degrees of freedom is responsible
for several fascinating properties of NEs. For example,
temperature change or illumination can change the
orientational order and cause the elastomer to extend or
contract by several hundred
percent~\cite{kuepfer_finkelmann_94,FinWer00,finkelmann_etal_01}.
Nematic elastomers display a soft
elasticity~\cite{golubovic_lubensky_89,FinKun97,VerWar96,War99}
characterized by vanishing shear stresses for a range of
longitudinal strains applied perpendicular to the nematic
direction. They exhibit an anomalous elasticity in which
certain bending and shear moduli are length-scale dependent
and vanish or diverge at long
length-scales~\cite{stenull_lubensky_epl2003,Xing_Radz_03,stenull_lubensky_anomalousNE_2003}.

In addition to unusual static properties, NEs have an
intriguing dynamic mechanical behavior. Early  rheology
experiments on liquid crystalline
elastomers~\cite{Gallani&Co} found no nematic effects in
NEs. Later, however, several
experiments~\cite{clarke&C0_2001,clarke&C0_damping_2001,stein&Co_2001,hotta_terentjev_2003}
observed a genuinely unconventional response to oscillatory
shear. It turns out that an internal relaxation of the
nematic director leads to a dynamic mechanical softening of
NEs. This behavior has been christened dynamic soft
elasticity. It makes NEs interesting for device
applications in areas such as mechanical vibration
damping~\cite{clarke&C0_damping_2001} (exploiting the fact
that the mechanical loss is record-high over a wide range
of temperatures and frequencies) or acoustics where NEs
open the possibility of acoustic
polarization~\cite{terentjev&Co_NEhydrodyn} (using the fact
that only particular soft shears are strongly attenuated)
analogous to the optical polarization in birefringent
media.

The theoretical investigation of the dynamic-mechanical
properties of NEs was pioneered by Terentjev, Warner and
coworkers
(TW)~\cite{WarnerTer2003,teixeira_warner_99,clarke&C0_2001,terentjev&Co_NEhydrodyn,terentjev_warner_2002}.
References~\cite{WarnerTer2003}
and~\cite{terentjev&Co_NEhydrodyn} present a detailed
derivation of equations governing the long-wavelength
low-frequency dynamics of NEs and derive their associated
mode structure, which includes non-hydrodynamic, rapidly
decaying modes. This derivation, which assumes a single
relaxation time for the director, is based on the
Lagrangian approach~\cite{goldstein} to the dynamics of
continuum systems in which non-dissipative forces are
calculated as derivatives of an energy functional and
dissipative forces as derivatives of a a Rayleigh
dissipation function. The TW work focuses primarily on
rheological response in both soft and semi-soft NEs at zero
wavenumber, and it neglects contributions to dynamical
equations arising from the Frank free energy for director
distortions, which are higher order in wavenumber than
those arising from network elasticity.  As a result, as we
shall see, the mode structure derived in
Refs.~\cite{WarnerTer2003} and
\cite{terentjev&Co_NEhydrodyn} misses diffusive modes along
symmetry directions in soft NEs.  When contributions from
the Frank free energy are included and those from the
elastic network are excluded, the TW approach reproduces
the original Leslie-Ericksen equations
\cite{ericksen_60,leslie_66} of nematodynamics, excluding
the rotational inertial term that is usually discarded
\cite{OrsayLC} or is missing entirely in alternative
derivations \cite{deGennesProst93} of these equations.

In this paper, we will present alternative approaches to
deriving the equations governing the dynamics of NEs. First
we will derive the exclusively hydrodynamical equations for
those variables whose characteristic frequencies vanish
with wavenumber.  These equations, like those for the
hydrodynamics of smectic
\cite{MartinPer1972,deGennesProst93} and columnar
\cite{deGennesProst93} liquid crystals, exclude
non-hydrodynamic director modes. Interestingly, the modes
for soft NEs predicted by these equations are identical in
form to those of columnar liquid crystals
\cite{deGennesProst93} with three pairs of sound modes and
diffusive modes along symmetry directions where sound
velocities vanish.  One pair of sound modes is
predominantly longitudinal with non-vanishing velocity at
all angles. A second pair is predominantly transverse with
a velocity that vanishes for wavevector either parallel or
perpendicular to the uniaxial symmetry axis, and a third
pair is completely transverse with a velocity that vanishes
for wavevector parallel to the symmetry axis but is nonzero
otherwise. We then derive the phenomenological equations
for the slow dynamics of all displacement and director
variables in NEs using the Poisson-Bracket formalism
\cite{PB,tomsBook} for obtaining the dynamics of
coarse-grained variables, applied so successfully to the
study of dynamical critical phenomena
\cite{HohenbergHal1977}, for deriving phenomenological
equations for any set of coarse-grained variables whose
dynamics is slow on a time scale set by microscopic
collision times. Like TW, we assume a single relaxation
time for the director.  We do, however, discuss how this
constraint can be relaxed.  When applied to fluid nematic
liquid crystals, the Poisson-bracket formalism is
equivalent to that used by the ``Harvard group"
\cite{forster&Co_71,Forster1983} in its derivation of
hydrodynamics of nematics.  Our dynamical equations for NEs
reduce to the equations of nematohydrodynamics derived by
the Harvard group when elastic rigidities vanish and to the
purely hydrodynamics equations for NEs when fast modes are
removed.  When contributions from the Frank free energy are
ignored, our equations are identical to those derived by
TW. We will not discuss random stresses or inhomogeneities
in this paper, though they contribute static components in
light scattering experiments \cite{gelmodes} that may
obscure the observation of the NE modes we discuss here.

The outline of our paper is as follows. In
Sec.~\ref{HarvardBasics} we briefly review the general
Poisson-bracket formalism for obtaining coarse-grained
dynamics. Section~\ref{pureHydrodynamics} focuses on the
pure hydrodynamics of NEs. We set up equations of motion
for the momentum density and the elastic displacement. We
derive the appropriate elastic energy entering these
equations by integrating out the director degrees of
freedom. Then we compare the hydrodynamic equations to
those of conventional uniaxial solids and columnar liquid
crystals. After extracting the sound velocities of the
modes we finally determine the full mode structure in the
incompressible limit. Section~\ref{low-frequencyDynamics}
features our formulation that explicitly accounts for the
dynamics of the director. We set up equations of motion for
the momentum density, the elastic displacement and the
director. These are then compared to the equations of
motion for uniaxial solids and to the equations of motion
derived by TW. We determine the mode structure. Finally, we
compare the so obtained modes to the modes obtained by TW.
A brief  summary is given in Sec.~\ref{concludingRemarks}.

\section{Coarse-grained Dynamics}
\label{HarvardBasics} Stochastic dynamical equations for
coarse-grained fields~\cite{HohenbergHal1977,tomsBook} can
be obtained by combining the Poisson-Bracket formalisms of
Classical mechanics \cite{goldstein}, which guarantees the
correct reactive couplings between fields with opposite
signs under time reversal, and the Langevin \cite{Langevin}
approach to stochastic dynamics, which provides a
description of dissipative processes and noise forces.  Let
$\Phi_\mu ( \brm{x} ,t )$, $\mu = 1, 2, ...$ be a set of
coarse-grained fields whose statistical mechanics is
described by a coarse-grained Hamiltonian $\mathcal{H}$.
The dynamical equations for $\Phi_\mu ( \brm{x},t)$ are
first-order differential equations in time:
\begin{align}
\label{genHarvard} &\dot{\Phi}_\mu (\brm{x}, t)
\nonumber \\
&= - \int d^d x^\prime \int d t^\prime \, \{ \Phi_\mu
(\brm{x}, t), \Phi_\nu (\brm{x}^\prime, t^\prime) \} \,
\frac{\delta \mathcal{H}}{\delta \Phi_\nu (\brm{x}^\prime,
t^\prime)}
\nonumber \\
& - \Gamma_{\mu, \nu}\, \frac{\delta \mathcal{H}}{\delta
\Phi_\nu (\brm{x}, t)} + \zeta_\mu (\brm{x}, t) \, .
\end{align}
Here and in the following the Einstein summation convention
is understood. The first term on the right hand side is a
non-dissipative velocity, also known as the reactive term,
that contains the Poisson bracket  $\{ \Phi_\mu (\brm{x},
t), \Phi_\nu (\brm{x}^\prime, t^\prime) \}$ of the coarse
grained fields~\cite{footnote1}. The reactive term couples
$\dot{\Phi}_\mu$ to $\delta \mathcal{H}/\delta \Phi_\nu$
only if $\Phi_\mu$ and $\Phi_\nu$ have opposite signs under
time reversal (when external magnetic fields are zero). The
second term on the right hand side is a dissipative term.
$\Gamma_{\mu, \nu}$ are the components of the so-called
dissipative tensor. This tensor couples $\dot{\Phi}_\mu$ to
$\delta \mathcal{H}/\delta \Phi_\nu$ only if $\Phi_\mu$ and
$\Phi_\nu$ have the same sign under time reversal. If the
noise term $\zeta_\mu$ is present, Eq.~(\ref{genHarvard})
represents a stochastic or Langevin equation. As such
Eq.~(\ref{genHarvard}) may be used to set up a dynamic
functional~\cite{Ja76,DeDo76,Ja92} to study the effects of
nonlinearities and fluctuations via dynamical field theory.
In this paper we are not interested in these effects.
Hence, we focus on linearized hydrodynamic equations and
pay no further attention to noise.

\section{Pure Hydrodynamics}
\label{pureHydrodynamics}

Hydrodynamics describes the dynamics of those degrees of
freedom whose characteristic frequencies $\omega$ vanish as
wavenumber tends to zero. In other words hydrodynamics
focuses exclusively on the {\em leading} low-frequency,
long-wavelength behavior. All other degrees of freedom,
even though they might be slow, are strictly speaking not
hydrodynamic ones. In this section we will derive the
hydrodynamic equations of NEs. These equations apply for
frequencies $\omega$ such that $\omega \tau \ll 1$, where
$\tau$ is the longest non-hydrodynamic decay time in the
system.  As we will show in the next section, the
characteristic time for director decay is in fact very slow
with $\tau \sim 10^{-2}$
sec.~\cite{schmidtke&CO_2000,schornstein&Co_2001} so that
the regime of applicability of hydrodynamics is quite small
for current NEs.  It is imaginable, however, that other
systems will be found with shorter decay times.

There are two general classes of hydrodynamic variables:
conserved variables and broken symmetry variables.  A
single-component NE has the same set of conserved variables
as a fluid: energy density $\epsilon$ , mass density
$\rho$, and momentum density $\brm{g}$. It also has the
same broken symmetry variables as a crystalline solid,
namely three displacement variables, though strictly
speaking, these variables in an elastomer are not
associated with a macroscopic broken symmetry because in
the absence of orientational order elastomers have the same
macroscopic rotational and translational symmetry as a
fluid. Since it is rotational symmetry that distinguishes a
nematic elastomer from an isotropic one, it could be argued
that the nematic director should be a hydrodynamic
variable, but as in smectic and columnar liquid crystals,
the director degrees of freedom decay in microscopic times
to their preferred configuration in the presence of strain
and are thus not hydrodynamic variables. Elastomers differ
from equilibrium crystals in at least two important ways:
The first, alluded to above, is that they are not periodic
and thus do not have mass-density wave order parameters
whose phases act as broken symmetry hydrodynamic variables;
rather Lagrangian displacement variables $\brm{u}$ take
their place. The second is that an elastomer is permanently
crosslinked: it is a classical rubber in which changes
$\delta \rho$ in mass density are locked to changes in
volume such that $\delta \rho/\rho = - {\bm \nabla} \cdot
\brm{u}$ and in which permeation in which there is
translation of a mass-density-wave without mass motion is
prohibited.  Thus, mass density is not an independent
hydrodynamic variable, and we are left with a total of
$5+3-1=7$ independent variables and $7$ associated
hydrodynamic modes.  These modes are heat diffusion and,
depending on direction, either six propagating sound modes,
four propagating sound modes and two diffusive
displacement/velocity modes, or two longitudinal sound
modes and diffusive displacement/velocity modes. In what
follows, we will consider only isothermal processes so that
heat heat diffusion can be ignored.

\subsection{Elastic energy}

To derive the hydrodynamical equations for NEs, we first
need the appropriate elastic free energy.  The elastic
constant measuring the energy of strains in planes
containing the anisotropy axis [the shear modulus $C_5$ to
be defined in Eq.~(\ref{uniax})] vanishes as a result of
the broken symmetry brought about by the establishment of
nematic order.  Thus a good starting point for this free
energy is that of a uniaxial solid with this elastic
constant simply set to zero.  This leaves certain soft
directions in which distortions cost zero energy, and as in
smectic and columnar liquid
crystals~\cite{deGennesProst93}, curvature-like terms that
are quadratic in second-order spatial derivatives have to
be added to ensure stability. To make contact with
dynamical equations involving the director to be presented
in Sec.~\ref{low-frequencyDynamics}, rather than simply
adding these terms, we find it useful to derive them from
the free energy of a nematic elastomer expressed in terms
of both the strain and the directors. We will restrict our
attention to harmonic distortions.

Elastomers are permanently crosslinked pieces of rubber
whose static elasticity is most easily described in
Lagrangian coordinates in which $\brm{x}$ labels a mass
point in the unstretched (reference) material and
$\brm{R}(\brm{x})= \brm{x} + \brm{u}(\brm{x})$, where
$\brm{u}(\brm{x})$ is the displacement variable, labels the
position of the mass point $\brm{x}$ in the stretched
(target) material.  We will use Lagrangian coordinates
throughout this paper.  We will, however on occasion make
reference to Eulerian coordinates in which $\brm{r}\equiv
\brm{R}(\brm{x})$ specifies a position in space and
$\brm{u}(\brm{r})$ the displacement variable at that
position.

The elastic energy of a nematic elastomer can be divided
into three parts:
\begin{eqnarray}
\label{fullEnergy} \mathcal{H} = \mathcal{H}_{\brm{u}} +
\mathcal{H}_{\brm{n}} + \mathcal{H}_{\brm{u},\brm{n}} \, ,
\end{eqnarray}
where $\mathcal{H}_{\brm{u}}$ is the usual elastic energy
of a uniaxial solid, $\mathcal{H}_{\brm{n}}$ is the Frank
free energy of a nematic, and
$\mathcal{H}_{\brm{u},\brm{n}}$ is the energy of coupling
between strain and director distortions.

Choosing the coordinate system so that the $z$-direction
coincides with the uniaxial direction, we have to harmonic
order \cite{LubenskyXin2002}
\begin{align}
\label{uniax} \mathcal{H}_{\brm{u}}& = \int d^3 x \,
\bigg\{ \frac{C_1}{2} \, u^2_{zz} + C_2 \, u_{zz} u_{ii} +
\frac{C_3}{2} \, u_{ii}^2
\nonumber \\
& + C_4 \, u_{ab}^2 + C_5 \, u_{az}^2 \bigg\} \, .
\end{align}
Here, $u_{ij}$ are linearized components of the Lagrange
strain tensor $\tens{u}$:
\begin{eqnarray}
u_{ij} = \frac{1}{2} \, \left( \partial_i u_j + \partial_j
u_i \right) \, .
\end{eqnarray}
We will use the convention that indices from the beginning
of the alphabet, $\{a,b\}$, assume the values 1 and 2
whereas indices from the middle of the alphabet,
$\{i,j,k,l\}$, run from 1 to 3. Note that, compared to our
work on the anomalous elasticity of
NEs~\cite{stenull_lubensky_epl2003,stenull_lubensky_anomalousNE_2003}
the elastic constants in Eq.~(\ref{uniax}) have a somewhat
different definition that is geared towards taking the
incompressible limit. Here we have arranged things so that
the terms featuring $C_2$ and $C_3$ involve the trace of
the strain tensor. In the incompressible limit that we will
eventually take one has $u_{ii} =0$ so that $C_2$ and $C_3$
drop out.

Expanded to harmonic order in the deviation $\delta \brm{n}
= \brm{n} - \brm{n}_0 $ from the uniform equilibrium state
$\brm{n}_0 = \hat{e}_z$, the Frank energy reads
\begin{align}
\label{frank} \mathcal{H}_{\delta \brm{n}}& = \int d^3 x \,
\bigg\{ \frac{K_1}{2} \,  (\partial_a \delta n_a)^2  +
\frac{K_2}{2}  \, (\varepsilon_{ab} \partial_a \delta
n_b)^2
\nonumber \\
&+ \frac{K_3}{2} \, (\partial_z \delta n_a)^2  \bigg\} \, .
\end{align}
where $\varepsilon_{ab}=-\varepsilon_{ba}$ is the
two-dimensional Levi-Civita symbol.  For notational
simplicity in the remainder of this paper we will replace
$\delta \brm{n}$ by $\brm{n}$ with the understanding that
it has only two components $n_a$.  With this notation,
$\mathcal{H}_{\delta \brm{n}}$ can be expressed as
\begin{equation}
\mathcal{H}_{\delta \brm{n}} = -\frac{1}{2} \int d^3 x \,
n_a M_{ab} (\brm{\nabla}) n_b ,
\end{equation}
where
\begin{equation}
M_{ab} = (K_1- K_2) \partial_a \partial_b + (K_2
\partial_{\perp}^2 + K_3 \partial_z^2 ) \delta_{ab}
\end{equation}
and where $\partial_{\perp}^2 = \partial_a \partial_a$ with
the Einstein convention understood.

The coupling energy, finally, can be written in the form
\begin{align}
\label{HunSimple} \mathcal{H}_{\brm{u},\brm{n}}& = \int d^3
x \, \bigg\{ \frac{D_1}{2} \,  Q_a^2  + D_2  \, u_{za} Q_a
\bigg\} \, ,
\end{align}
where once more terms beyond harmonic order have been
neglected and where
\begin{eqnarray}
\label{defQ} Q_a = \delta n_a - \frac{1}{2} \, (\partial_z
u_a -
\partial_a u_z) \equiv \delta n_a - \tilde{\Omega}_a \, .
\end{eqnarray}

As explained above the director is not a genuine
hydrodynamic variable and hence it should be integrated out
of the elastic energy as long as we focus on the
hydrodynamic limit. We do so by minimizing $\mathcal{H}$
over $n_a$ to find
\begin{equation}
n_a = \tilde{\Omega}_a - \frac{D_2}{D_1} u_{az} +
\frac{M_{ab} ( \brm{\nabla} )}{D_1} \, \left(
\tilde{\Omega}_b - \frac{D_2}{D_1} u_{bz} \right) .
\end{equation}
Inserting this equation into $\mathcal{H}$ and retaining
only the dominant terms in gradients, we obtain
\begin{align}
\label{H-u-el} \mathcal{H}_{\brm{u}-{\rm el}} & = \int d^3
x \, \bigg\{ \frac{C_1}{2} \, u^2_{zz} + C_2 \, u_{zz}
u_{ii} + \frac{C_3}{2} \, u_{ii}^2
\nonumber \\
& + C_4 \, u_{ab}^2 + C_5^R \, u_{az}^2 + \frac{K_1^R}{2}
(\partial_{\perp}^2 u_z )^2  + \frac{K_3^R}{2}
(\partial_z^2 u_a)^2 \bigg\} \, .
\end{align}
where
\begin{subequations}
\label{reno}
\begin{align}
\label{softnessCond} C_5^R & = C_5 - \frac{D_2^2}{2 D_1} \,
,
\\
K_1^R & = \frac{1}{4} \left(1 + \frac{D_2}{D_1}\right)^2
K_1 \, ,
\\
K_3^R & = \frac{1}{4} \left(1- \frac{D_2}{D_1}\right)^2 K_3
\, .
\end{align}
\end{subequations}
The superscript $R$ indicates that $C_5$, $K_1$ and $K_3$
have been renormalized by director fluctuations. Note that
the elastic constants $K_1^R$ and $K_3^R$ are not the same
as the Frank splay and bend constants $K_1$ and $K_3$. This
is in contrast to smectic-A and columnar liquid crystals in
which the coefficients of quartic gradient terms in the
effective elastic energy arising from the relaxation of
director modes are identical to the Frank elastic
constants. The modulus $D_2$ can have either sign, and
there are no thermodynamic constraints preventing either
$1+D_2/D_1$ or $1-D_2/D_1$ from being zero. Thus, either
$K_1^R$ or $K_3^R$ in Eq.~(\ref{reno}) could be zero. In
this case, contributions to $(\partial_{\perp}^2 u_z )^2$
or $(\partial_z^2 u_a)^2$ in $\mathcal{H}_{\brm{u}-{\rm
el}}$ arising from network elasticity would have to be
added to ensure stability. Except for the exceptional cases
when $D_2 = \pm D_1$, those contributions are generally
smaller than the ones represented in Eq.~(\ref{reno}), and
we will ignore them. If $C_5^R$ vanishes, then the
dependence of the elastic energy on $u_{az}$ drops out.
This is the famous soft elasticity. If the condition for
softness~(\ref{softnessCond}) is not strictly fulfilled and
there is a small but finite remnant $C_5^R$ then small
shears in the planes containing the anisotropy axis do cost
a small energy. This non-ideal behavior is known as
semi-softness.

The energy~(\ref{H-u-el}) combines attributes of both
smectic and columnar liquid crystals.  The vertical
displacement $u_z$ is analogous to the displacement
variable $u$ of a smectic liquid crystal.  It is soft for
distortions in the $\perp$-direction, and a bending term
$K_1^R (\partial_{\perp}^2 u_z)^2/2$ is needed to stabilize
it.  The in-plane displacements $u_a$ are analogous to
those of columnar liquid crystal, which are soft for
distortions in the $z$-direction, and a bending term $K_3^R
(\partial_z^2 u_a)^2/2$ is need to stabilize them.

\subsection{Hydrodynamic equations}

We are now in a position to write down the full
hydrodynamic equations for nematic elastomers.  For
simplicity, we will restrict our attention to isothermal
processes so that we can ignore temperature diffusion. This
leaves us with six hydrodynamical variables, the momentum
density $g_i(\brm{x})$ and the displacement $u_i(\brm{x})$.
These are independent variables at the reference point
$\brm{x}$ that satisfy the continuum generalizations of the
usual relations for the momentum and displacement of a
particle: $\partial g_i (\brm{x})/\partial g_j (\brm{x}') =
\delta_{ij} \delta ( \brm{x} - \brm{x}')$, $\partial u_i
(\brm{x})/\partial u_j (\brm{x}') = \delta_{ij} \delta (
\brm{x} - \brm{x}')$, and $\partial g_i (\brm{x})/\partial
u_j (\brm{x}') = 0$.  These relations yield a non-vanishing
Poisson bracket between $u_i ( \brm{x})$ and
$g_j(\brm{x}')$
\begin{equation}
\{u_i ( \brm{x}),g_j(\brm{x}')\} = \delta_{ij}
\delta(\brm{x} - \brm{x}') .
\end{equation}
The $\brm{g}-\brm{g}$ and $\brm{u}-\brm{u}$ Poisson
brackets are zero. Using these results, we obtain the
equations of motion
\begin{subequations}
\label{genStructSimple}
\begin{align}
\label{EOMsimple1a} v_i & \equiv \dot{u}_i  = \frac{\delta
\mathcal{H}_{\rm kin}}{\delta g_i} = \frac{1}{\rho} \, g_i
\, ,
\\
\label{EOMsimple1b}
 \dot{g}_i &=   - \frac{\delta
\mathcal{H}_{\brm{u}-\text{el}}}{\delta u_i} +
\eta_{ijkl}\, \partial_j
\partial_l v_k\,  ,
\end{align}
\end{subequations}
where $\brm{v}$ is the velocity field, and
$\mathcal{H}_{\rm kin} = \int d^3 x g_i g_i/(2 \rho)$ is
the coarse-grained kinetic energy.  $\eta_{ijkl}$ is the
viscosity tensor, which has five independent components. We
can parametrize the stress tensor so that the entropy
production from viscous stresses takes on the same form as
the elastic energy $\mathcal{H}_{\brm{u}}$,
\begin{align}
\label{uniaxViscosity} T  \dot{S}& = \int d^3 x \, \bigg\{
\frac{\eta_1}{2} \, \dot{u}^2_{zz} + \eta_2 \, \dot{u}_{zz}
\dot{u}_{ii} + \frac{\eta_3}{2} \, \dot{u}_{ii}^2
\nonumber \\
& + \eta_4 \, \dot{u}_{ab}^2 + \eta_5^R \, \dot{u}_{az}^2
\bigg\} \, ,
\end{align}
where the $\eta$'s are low frequency viscosities.
$\eta_5^R$ is an effective viscosity that is, as its
counterpart $C_5^R$, renormalized by director fluctuations.
Equation (\ref{uniaxViscosity}) contains all contributions
to the entropy production equation in the truly
hydrodynamic limit we are considering. Nonhydrodynamic
variables like the director have already relaxed to their
local equilibrium values. Thermodynamic stability requires
$\eta_i \geq 0$ for $i = 1,3,4$, $\eta_5^R\geq 0$ and
$\eta_1 \eta_3 \geq \eta_2^2$.

A few observations about Eqs.~(\ref{genStructSimple}) are
in order.  First they are identical in form to the
equations for a conventional elastic material (without
vacancy diffusion). The distinction between such a material
and a nematic elastomer appears only in the form of
$\mathcal{H}_{\brm{u}-{\rm el}}$.  The absence of any
dissipative term proportional to $-\delta
\mathcal{H}/\delta u_i$ in Eq.~(3.13a) reflects the
tethered or crosslinked character of the elastomer. In
non-crosslinked systems, this equation would contain an
additional dissipative term proportional to $- \delta
{\mathcal{H}}_{\brm{u}-\text{el}}/\delta u_i$ describing
permeation. Second, these equations are expressed in
Lagrangian coordinates, and the derivatives $\dot{\brm{u}}$
and $\dot{\brm{g}}$ are time derivatives at constant value
of the reference position $\brm{x}$. In Eulerian
coordinates, $\brm{u}$ and $\brm{g}$ become functions of
points $\brm{r}= \brm{R}(\brm{x})$ in space:
$\brm{u}_E(\brm{r},t) = \brm{u}(\brm{x}(\brm{r}),t)$ and
similarly for $\brm{g}_E$. The time derivative
$\dot{\brm{u}}=d \brm{u}/dt =
\partial \brm{u}_E /\partial t + \brm{v} \cdot \brm{\nabla}
\, \brm{u}_E$ is the time derivative in the reference frame
moving with the local fluid velocity.

To obtain a more explicit form for the equations of motion,
we note that Eq.\ (\ref{EOMsimple1a}) allows us to replace
${\dot g}_i$ in Eq.\ (\ref{EOMsimple1b}) with $\rho
\ddot{u}_i$ to produce the standard mechanical equation for
a solid with dissipation,
\begin{subequations}
\label{finalSimpleEOM}
\begin{align}
\rho \ddot{u}_a & =  (C_2 + \eta_2 \partial_t) \partial_a
u_{zz} + (C_3 + \eta_3 \partial_t) \partial_a u_{ii}
\nonumber\\
&+ 2\, (C_4 + \eta_4 \partial_t) \partial_b u_{ab} + (C_5^R
+ \eta_5^R \partial_t) \partial_z u_{az}
\nonumber\\
&- K_3^R \partial_z^4 u_a\, ,
\\
\rho \ddot{u}_z & = (C_1 + \eta_1 \partial_t + C_2 + \eta_2
\partial_t) \partial_z u_{zz}
\nonumber\\
&+ (C_2 + \eta_2 \partial_t + C_3 + \eta_3 \partial_t)
\partial_z u_{ii}
\nonumber \\
&+ (C_5^R + \eta_5^R \partial_t) \partial_b u_{bz} - K_1^R
\partial_\perp^4 u_z \, .
\end{align}
\end{subequations}
Apart from the appearance of $C_5^R$, which is zero in a
soft NE, rather than $C_5$ and bending terms with quartic
derivatives, these equations are identical to those of a
uniaxial solid.

Upon switching to frequency space one can see that the
static renormalization of the shear modulus $C_5$,
Eq.~(\ref{softnessCond}), has a dynamical generalization
with a frequency dependent $C_5^R (\omega)= C_5^R - i
\omega \eta_5^R$, whose real part vanishes when $C_5^R =
0$. This form for $C_5^R( \omega)$ is only valid in the
hydrodynamic limit with $\omega \tau \ll 1$.  As we will
discuss more fully in the next section, there are other
terms in $C_5^R ( \omega)$ when this limit is not obeyed.

\subsection{Sound velocities}
To assess the mode structure of the equations of
motion~(\ref{finalSimpleEOM}), we begin with an analysis of
propagating sound modes in the absence of dissipation. To
keep arguments simple, we will restrict ourselves here to
ideally soft NEs with $C_5^R=0$. When $C_5^R \neq 0$, the
bending terms can be neglected, and the dynamical equations
and associated equations are identical to those of uniaxial
solids. We will return to the case when $C_5^R$ is nonzero
in Sec.~\ref{low-frequencyDynamics}.

\begin{figure}
\includegraphics[width=5.0cm]{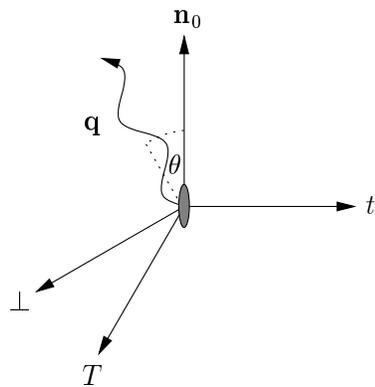}
\caption[]{\label{coordinates}Symmetry directions. The
$t$-direction is perpendicular to the plane containing the
equilibrium director $\brm{n}_0$ and the wavevector
$\brm{q}$ of a generic excitation. The $T$-direction is
perpendicular to $\brm{q}$ and the $t$-direction.}
\end{figure}

The sound modes have frequencies $\omega(\brm{q}) =
c(\theta) \, q$ where $\theta$ is the angle that $\brm{q}$
makes with the $z$-axis.  The non-dissipative bending terms
give rise to modes with $\omega \sim q^2$ along the
symmetry direction.  These modes, however, mix with
dissipative ones and become overdamped diffusive modes with
$\omega \sim - i q^2$. Thus, to obtain the true
non-dissipative sound-mode structure, we can ignore the
bending terms. Consequentially, the non-dissipative
sound-mode structure is that of a uniaxial solid with $C_5
= 0$, which it turns out is identical to that of columnar
liquid crystals, though the input variables are slightly
different (there is no $u_z$ in a columnar liquid crystal,
but there is an independent density).  The modes break up
into displacements $u_t = \varepsilon_{ab} q_a
u_b/q_{\perp}$ perpendicular to both $\brm{q}$ and the
$z$-axis and coupled displacements $u_{\perp} = q_a
u_a/q_{\perp}$ and $u_z$ in the plane that contains
$\brm{q}$ (see Fig.~\ref{coordinates}).  In Fourier space,
the hydrodynamic equations are
\begin{subequations}
\begin{align}
\rho \omega^2 u_t & =  C_4 \, q_{\perp}^2 u_t \, ,
\\
\rho \omega^2 u_{\perp} & =  (C_2 + C_3) \, q_\perp q_z u_z
+ (C_3 + 2C_4) \, q_\perp^2  u_\perp \, ,
\\
\rho \omega^2 u_z & =  (C_1 + 2C_2 + C_3)  \, q_z^2 u_z +
(C_2 + C_3) \, q_\perp q_z  u_\perp \, .
\end{align}
\end{subequations}

\begin{figure}
\includegraphics[width=8.0cm]{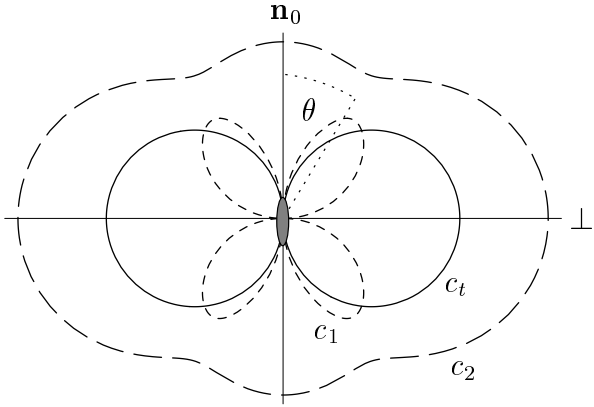}
\caption[]{\label{soundvelocities} Schematic polar plot
(arbitrary units) of the sound velocities $c_t$ (solid
line), $c_1$ (short dashes), and $c_2$ (long dashes).
}
\end{figure}

Thus, there is a transverse sound mode with velocity
\begin{subequations}
\begin{equation}
c_t(\theta) =\sqrt{\frac{C_4}{\rho}} | \sin\theta |
\end{equation}
that vanishes when $\theta=0$ and reaches a maximum at
$\theta = \pi/2$.  The sound velocities $c_1(\theta)$ and
$c_2(\theta)$ of the other modes are coupled and satisfy
\begin{align}
c_1^2\,  c_2^2 &= \frac{1}{\rho^2} [(2 C_4 + C_3)(C_1+2 C_2
+
C_3) \nonumber \\
& - (C_2+C_3)^2]\cos^2 \theta \sin^2 \theta ] \\
c_1^2 + c_2^2 &=\frac{1}{\rho}[(2 C_4 + C_3) \sin^2 \theta
+ (C_1 + 2C_2 + C_3) \cos^2 \theta ]
\end{align}
\end{subequations}
These equations are identical to the equations satisfied by
the sound velocities in a columnar
system~\cite{deGennesProst93}.  One of the sound modes is
purely longitudinal in the limit $C_3 \to \infty$; its
velocity  is non-vanishing for all $\theta$. The second
mode is like that of a smectic-$A$
\cite{deGennesProst93,MartinPer1972} with a sound velocity
that vanishes at $\theta=0$ and $\theta=\pi/2$. The third
mode is a purely transverse sound mode whose velocity
vanishes only at $\theta =0$. Figure~\ref{soundvelocities}
plots the three sound velocities.

\subsection{Full incompressible mode structure}
\label{fullModeStructure} Having found the general sound
mode structure in the non-dissipative limit, we turn to a
full analysis of modes in the incompressible limit.  To
discuss this limit, it is useful to decompose
$\brm{u}(\brm{q})$ into a longitudinal part $u_l$ along
$\brm{q}$ and components $u_t$, perpendicular to both
$\brm{q}$ and $z$, and $u_T$ perpendicular to $\brm{q}$ in
the plane containing $\brm{q}$ and $z$ as shown in
Fig.~\ref{coordinates}.

As was the case in the dissipationless limit, $u_t$
decouples from $u_T$ and $u_l$. In the incompressible limit
$u_l$ vanishes and we are left with
\begin{subequations}
\begin{align}
\rho \omega^2 u_t & = (C_4 q_{\perp}^2 + K_3^R q_z^4) u_t -
i \omega (\eta_4 q_{\perp}^2 + \eta_5^R q_z^2 ) u_t \, ,
\\
\rho \omega^2 u_T & = \left[ (C_1 + 2 C_4)
\frac{q_{\perp}^2 q_z^2}{q^2} + K_1^R \,
\frac{q_{\perp}^6}{q^2} + K_3^R\, \frac{q_z^6}{q^2} \right]
\, u_T
 \nonumber \\
& - i \omega \left[(\eta_1 + 2 \eta_4)\frac{q_{\perp}^2
q_z^2}{q^2}+ \frac{\eta_5^R}{2} \, \frac{(q_{\perp}^2-
q_z^2)^2}{q^2} \right] \, u_T \, .
\end{align}
\end{subequations}
These equations produce propagating modes with respective
frequencies
\begin{subequations}
\label{propModes}
\begin{align}
\label{fullt} \omega_{t,\pm} & = \pm\sqrt{\frac{C_4}{\rho}}
\,  |q_{\perp}| - i\, \frac{2\eta_4 \, q_{\perp}^2 +
\eta_5^R \, q_z^2}{4\, \rho} \, ,
\\
\label{fullT} \omega_{T,\pm} & = \pm\sqrt{\frac{C_1 +
2C_4}{\rho}} \, \frac{|q_{\perp} q_z|}{q}
\nonumber \\
&- i\, \frac{2(\eta_1 + 2\eta_4) \, q_\perp^2 q_z^2 +
\eta_5^R \, (q_\perp^2 - q_z^2)^2}{4 \,  \rho \, q^2} \, ,
\end{align}
\end{subequations}
as long as $\brm{q}$ is not along a symmetry direction in
which the sound velocity is zero.  When $q_{\perp}=0$, the
 $t$- and $T$-modes become diffusive with identical frequencies
\begin{equation}
\label{diffusiveModestT} \omega_{t,\pm} = \omega_{T,\pm} =
\frac{1}{4 \rho} \left[ -i \eta_5^R  \pm \sqrt{-
(\eta_5^R)^2  + 16 \rho K_3^R } \, \right] \, q_z^2 \, .
\end{equation}
As in conventional nematics $K_1\sim K_3 \sim 10^{-6} \
\text{dynes}$ and $\rho \sim 1\  \text{gram}/\text{cm}^3$.
The viscosity should be larger than the $0.01 \
\text{poise}$ characteristic of fluids. Thus, we can expect
that $16 \rho K_3^R \ll (\eta_5^R)^2$. In this limit the
above modes become slow and a fast diffusive mode with
frequencies
\begin{subequations}
\label{diffusiveModestTslowFast}
\begin{align}
\label{slowBoth}
\omega_{t,s} &= \omega_{T,s} = - i\, \frac{ 2
K_3^R}{\eta_5^R} q_z^2 \, ,
\\
\omega_{t,f} & = \omega_{T,f} = - i\, \frac{\eta_5^R}{2
\rho} q_z^2 \, .
\end{align}
\end{subequations}
When $q_z=0$, the $t$-modes remain propagating sound modes,
but the $T$-modes become diffusive with frequencies
\begin{equation}
\label{noIdea1} \omega_{T,\pm} = \frac{1}{4 \rho} \left[ -i
\eta_5^R  \pm \sqrt{- (\eta_5^R)^2  + 16 \rho K_1^R }\,
\right] \, q_{\perp}^2 \, ,
\end{equation}
which in the limit $16 K_1^R \rho \ll (\eta_5^R)^2$ reduce
to
\begin{subequations}
\label{noIdea2}
\begin{align}
\label{slowT}
\omega_{T,s} &= - i \, \frac{ 2 K_1^R}{\eta_5^R}
q_{\perp}^2\, ,
 \\
 \label{fastT}
\omega_{T,f} & = - i\, \frac{\eta_5^R}{2 \rho} q_{\perp}^2
\, .
\end{align}
\end{subequations}

Note that there is one diffusive and one propagating sound
mode in the incompressible limit when $q_z = 0$. The
attenuation (the imaginary part of q) of the diffusive
modes is proportional to $\sqrt{\omega}$ and that of the
sound modes is proportional to $\omega$. This is the
explanation of the large difference in attenuation of the
two polarizations of transverse wave found by TW
\cite{terentjev&Co_NEhydrodyn} that makes NEs candidates
for acoustic polarizers. We will return to this issue in
Sec.~\ref{compTW}.

Equations~(\ref{slowBoth}) and (\ref{slowT}) show that one
misses the slow diffusive modes if one neglects the Frank
energy. It is a legitimate question, however, under what
conditions these modes can be observed experimentally.
Manifestly, they should be observable directly in the
symmetry directions in which the respective sound
velocities vanish whereas they should not be seen in
directions which differ significantly from these symmetry
directions. Between the two extremes there will be a
crossover from slow diffusive to propagating behavior at
certain crossover angles. These angles can be estimated by
comparing the magnitude of the terms in
Eqs.~(\ref{propModes}) and (\ref{slowBoth}) or respectively
(\ref{slowT}). For $\theta \approx 0$ we get for example
from Eqs.~(\ref{propModes}) and (\ref{slowBoth}) that the
crossover is expected at an angle $\theta_0$ such that
\begin{align}
|\sin \theta_0 | \approx \sqrt{\frac{\rho}{C_4}}\,
\frac{K_3^R}{\eta_5^R} \, q \, .
\end{align}
The wave vectors in light scattering experiments on NE
dynamics typically have a magnitude $q \sim 10^5 \
\text{cm}$. The elastic moduli $C_1$ and $C_4$ should be
comparable to the shear modulus of rubber, $C_1\sim C_4
\sim 10^7 \ \text{dynes}/\text{cm}^2$. Hence, we estimate
\begin{align}
|\sin \theta_0 | \approx 10^{-4} \, ,
\end{align}
i.e., the range around the nematic direction in which the
slow diffusive behavior is observable is extremely narrow.
Applying the same reasoning to the slow diffusive $T$-mode
for $\theta \approx \pi/2$ we obtain a further crossover
angle $\theta_{\pi/2}$ with
\begin{align}
|\cos \theta_{\pi/2} | \approx 10^{-4} \, ,
\end{align}
signaling the same extremely narrow angle range for slow
diffusive behavior as above.

\section{Dynamics with displacements and director}
\label{low-frequencyDynamics}
Nematic elastomers, like their conventional
nematic-liquid-crystal counterparts, are characterized by a
Frank director $\brm{n}$ that responds dynamically to
external forces.  In this section, we will use the
Poisson-bracket approach to derive phenomenological
equations for the dynamics of both the director and
displacements in nematic elastomers.  Though our derivation
is different from that of TW, our dynamical equations are
in fact identical to theirs if they include contributions
from the Frank free energy for the director.  Our equations
also reduce to the standard equations of
nematohydrodynamics \cite{forster&Co_71,Forster1983} when
elasticity due to network crosslinking is turned off. It
must be emphasized, however, that our equations predict
non-hydroydynamic modes characterized by a decay time
$\tau$ that does not approach infinity with vanishing
wavenumber $q$. In any real system, there are many
non-hydrodynamic modes with characteristic decay times
$\tau_{\alpha}$.  For a dynamical theory to provide a
correct description of a system over a frequency range from
zero to some maximum frequency $\omega_M$, it must include
contributions from all modes, both hydrodynamic and
non-hydrodynamic, with characteristic frequencies up to a
few times $\omega_M$. An isotropic rubber is characterized
by rather long decay times $\tau_R$ of Rouse-like modes of
chain segments~\cite{rubinstein_colby_2003}. At frequencies
$\omega$ such that $\omega \tau_R \ll 1$, its viscosities
become frequency independent described by the
hydrodynamical equations of an isotropic solid.  However,
when $\omega \tau_R \geq 1$, viscosities develop nontrivial
frequency dependence, and hydrodynamics breaks down.  In
our theory for nematic elastomers, we assume that there is
a single director relaxation time $\tau$ and that it is
much larger than $\tau_R$ so that we do not need to worry
about the frequency dependence of viscosities arising from
Rouse modes. We will, however, point out where these
assumptions, also made by TW, can be modified.   If $\tau
\leq \tau_R$, dynamics at frequencies $\omega \tau_R \geq
1$ will be dominated by Rouse modes, and it may be
difficult to distinguish director relaxation modes from
Rouse modes in experiments in which displacements are
probed. The hydrodynamic description of the preceding
section, however, remains valid when $\omega \tau_R \ll 1$.

\subsection{Equations of motion}

When we keep the director as a dynamical variable, our
formulation is  closely related to nematodynamics. The
equation of motion for the director has a reactive coupling
to the velocity $\brm{v} = \dot{\brm{u}}$ arising from the
Poisson bracket of $n_i (\brm{x} )$ with $g_j (\brm{x}')$.
This Poisson bracket is generally derived in Eulerian
coordinates~\cite{Forster1983,stark_lubensky_03} in which
there is a contribution $\brm{v} \cdot \brm{\nabla}
\brm{n}$ to the equation for $\partial_t \brm{n}$. This
term can be combined with $\partial_t \brm{n}$ to yield the
Lagrangian time derivative $\dot{\brm{n}}$.  The remaining
part of the Poisson bracket is
\begin{equation}
\{n_i(\brm{x}), g_j (\brm{x}^\prime)\}= -\lambda_{ijk}
\partial_k \delta (\brm{x} - \brm{x}') .
\end{equation}
The properties of the tensor $\lambda_{ijk}$ are dictated
by three constraints: First, the magnitude of the director
has to be conserved, i.e., $\brm{n} \cdot \dot{\brm{n}} =0$
implying  $n_i \lambda_{ijk} = 0$; second, the equations of
motion must be invariant under $\brm{n} \rightarrow -
\brm{n}$ implying $\lambda_{ijk}$ must change sign with
$\brm{n}$; and third, under rigid uniform rotations, the
director has to obey $\dot{\brm{n}} = \frac{1}{2} (\nabla
\times \dot{\brm{u}}) \times \brm{n}$. The only tensors and
vectors available to use for the construction of
$\lambda_{ijk}$ are $\delta_{ij}$, $n_i$, and the
Levi-Civita anti-symmetric tensor $\epsilon_{ijk}$, and the
only tensors that satisfy the first two conditions above
are $\delta^T_{ij} n_k$ and $\delta^T_{ik} n_j$ where
$\delta_{ij}^T = \delta_{ij} - n_i n_j$.  Thus, the first
two conditions imply that $\lambda_{ijk}$ has two
independent components, which can be expressed as parts
symmetric and antisymmetric under interchange of $j$ and
$k$:
\begin{eqnarray}
\lambda_{ijk} = \frac{\lambda}{2} \, \left( \delta_{ij}^T
\, n_k  + \delta_{ik}^T \,n_k \right)+ \frac{\lambda_2}{2}
\, \left( \delta_{ij}^T \, n_k  - \delta_{ik}^T \,n_k
\right) ,
\end{eqnarray}
The third condition implies that the coefficient
$\lambda_2$ of the antisymmetric part must be equal to
$-1$. There are no constraints on the value of $\lambda$,
which is equal to the ratio of two dissipative coefficients
of the Leslie-Eriksen theory \cite{forster&Co_71}.  When
its absolute magnitude is positive, it determines the
equilibrium tilt angle of the director under uniform shear
\cite{leslie_66}.

Using the Poisson brackets for the director and those for
the momentum density and displacement discussed in the
preceding section, we obtain the equations of motion for a
nematic elastomer
\begin{subequations}
\label{genStruct}
\begin{eqnarray}
\label{EOMa} \dot{n}_i &=& \lambda_{ijk} \, \partial_j
\dot{u}_k - \Gamma \, \frac{\delta \mathcal{H}}{\delta n_i}
\, ,
\\
\label{EOMb} \dot{u}_i &=& \frac{1}{\rho} \, g_i \, ,
\\
\label{EOMc} \dot{g}_i &=& \lambda_{kji} \, \partial_j
\frac{\delta \mathcal{H}}{\delta n_k}  - \frac{\delta
\mathcal{H}}{\delta u_i} + \nu_{ijkl}\, \partial_j
\partial_l \dot{u}_k \,
.
\end{eqnarray}
\end{subequations}
The first terms on the respective right hand sides of
Eqs.~(\ref{EOMa}) and (\ref{EOMb}) as well as the first and
second term on the right hand side of Eq.~(\ref{EOMc}) stem
from the Poisson brackets. $\mathcal{H}$ is the full
elastic energy of NEs as stated in Eq.~(\ref{fullEnergy}).
The second term on the right hand side of Eq.~(\ref{EOMa})
is a dissipative term that describes diffusive relaxation.
There is no dissipative contribution to Eq.~(\ref{EOMb})
because NEs are tethered. $\nu_{ijkl}$ is a viscosity
tensor that has the same structure as $\eta_{ijkl}$. It has
five independent components $\nu_1$ to $\nu_5$. We use
different symbols here for the viscosities than we used in
Sec.~\ref{pureHydrodynamics} so that we can cleanly keep
track of differences between the two theories.

We have assumed in Eqs.~(\ref{genStruct}) that both
$\Gamma$ and $\nu_{ijkl}$ are local in time or,
equivalently, that their temporal Fourier transforms are
independent of frequency as they are at frequencies
$\omega$ less than their respective inverse characteristic
times $\tau_\Gamma^{-1}$ and $\tau_R^{-1}$. At larger frequencies, however, both $\Gamma$ and
$\nu_{ijkl}$ do depend on frequency.  In polymer gels,
viscosities are proportional to $\sqrt{\omega}$ at
frequencies $\omega \tau_R \gg 1$ because of the large
number of closely spaced modes contributing at these
frequencies. Though there are to our knowledge no
microscopic calculations of $\Gamma(\omega)$, there is no
reason why many closely spaced modes should not lead to
Rouse-like behavior at frequencies $\omega \tau_\Gamma \gg
1$. There is also no reason why $\tau_\Gamma$ and $\tau_R$
should not be equal or nearly so.  If the decay time $\tau$
for the nonhydrodynamic director modes predicted by the the
theory with the low frequency approximations to $\Gamma$
and $\nu_{ijkl}$ is greater than $\tau_R$ and
$\tau_\Gamma$, then this theory provides a correct
description of the dynamics for frequencies $0< \omega <
{\rm min}( \tau_R^{-1},\tau_\Gamma^{-1})$ including $\omega
\sim \tau^{-1}$.  If on the other hand $\tau < {\rm
min}(\tau_R, \tau_\Gamma)$, the full frequency dependence
of $\Gamma$ must be used for frequencies $\omega > {\rm
min}(\tau_R^{-1}, \tau_\Gamma^{-1})$.  In all cases,
however, the hydrodynamic theory of the preceding section
is regained in the limit $\omega \ll {\rm
min}(\tau^{-1},\tau_R^{-1},\tau_\Gamma^{-1})$. To keep our
discussion simple, we will continue to assume that both
$\Gamma$ and $\nu_{ijkl}$ are frequency independent. Our
equations can, however, incorporate the frequency
dependence of these quantities merely by replacing them by
their frequency-dependent forms, $\Gamma(\omega)$ and
$\nu_{ijkl}(\omega)$.

At this point we would like to comment on the precise
relation of Eqs.~(\ref{genStruct}) to the equations of
nematodynamics. One retrieves the latter from the former by
setting all the elastic constants in $\mathcal{H}$, except
for the Frank constants $K_1$, $K_2$, and $K_3$, equal to
zero. Thus, we are guaranteed to obtain the well known
modes of nematic liquid crystals in this limit. One can
also make the observation that in this limit
Eqs.~(\ref{genStruct}) depend just on $\brm{v} =
\dot{\brm{u}}$ rather than on $\brm{v}$ and $\brm{u}$. If
one seeks modes in terms of $\brm{u}$ rather than $\brm{v}$
one finds  extra zero frequency modes that are spurious.

Next we condense the equations of motion into a reduced set
of effective equations. We can simplify the subsequent
analysis at the onset by eliminating either $\brm{u}$ or
$\brm{g}$ with help of Eq.~(\ref{EOMb}). To facilitate in
contact with the work of TW we choose to keep $\brm{u}$.
For the same reason we opt to work with $\brm{Q}$ rather
than $\brm{n}$. This represents no difficulty since
$\brm{Q}$ and $\brm{n}$ are simply related via
Eq.~(\ref{defQ}). Collecting we obtain after some algebra
the following effective equations of motion for $\brm{Q}$
and $\brm{u}$:
\begin{subequations}
\label{effectiveEOMs}
\begin{align}
\label{effectiveEOMsa}
&\big\{ [\partial_t + \Gamma D_1 ] \delta_{ab} - \Gamma
M_{ab} (\brm{\nabla})\big\} \, Q_b
\nonumber\\
& = [\lambda \partial_t - \Gamma D_2 ] \, u_{az} + \Gamma
M_{ab} (\brm{\nabla}) \, \tilde{\Omega}_b \, ,
\end{align}
and
\begin{align}
\label{effectiveEOMsb}  \rho \partial_t^2 u_a &=  -
\frac{\lambda+1}{2\, \Gamma}\, \partial_t \partial_z [Q_a -
\lambda u_{az}] + [C_2 + \nu_2 \partial_t] \partial_a
u_{zz}
\nonumber\\
&+ [C_3 + \nu_3 \partial_t] \partial_a u_{ii} + 2 [C_4 +
\nu_4
\partial_t] \partial_b u_{ab}
\nonumber\\
&+ \left[C_5 + \nu_5 \partial_t - \frac{D_2}{2} \right]
\partial_z u_{az} +  \frac{D_2 - D_1}{2}\, \partial_z Q_a\,
,
\\
\label{effectiveEOMsc}  \rho \partial_t^2 u_z &=  -
\frac{\lambda-1}{2\, \Gamma}\, \partial_t \partial_b [Q_b -
\lambda u_{bz}]
\nonumber\\
&+ [C_1 + \nu_1 \partial_t + C_2 + \nu_2 \partial_t]
\partial_z u_{zz}
\nonumber\\
&+ [C_2 + \nu_2 \partial_t + C_3 + \nu_3 \partial_t]
\partial_z u_{ii}
\nonumber\\
&+ \left[C_5 + \nu_5 \partial_t + \frac{D_2}{2} \right]
\partial_b u_{bz} +  \frac{D_1 + D_2}{2}\, \partial_b Q_b\, .
\end{align}
\end{subequations}

To make contact to uniaxial solids we now take a brief
detour and consider the simplified case $M_{ab}
(\brm{\nabla}) =0$. In this case Eq.~(\ref{effectiveEOMsa})
is diagonal in frequency space and readily solved with the
result
\begin{align}
\label{simpleQa} Q_a = - \frac{D_2}{D_1} \, \frac{1- i
\omega \tau_2}{1 - i \omega \tau_1} \, u_{az} \, .
\end{align}
In writing Eq.~(\ref{simpleQa}) we have used the relaxation
times $\tau_1 = 1/(\Gamma D_1)$ and $\tau_2 =
-\lambda/(\Gamma D_2)$. $\tau_1 \approx
10^{-2}s$~\cite{WarnerTer2003,schmidtke&CO_2000,schornstein&Co_2001}
is essentially the relaxation time of the director.  Pure
hydrodynamic behavior is obtained when both $\omega \tau_1$
and $\omega \tau_2$ are much less than 1 as our
calculations will verify. In terms of the relaxation times,
$\lambda$ is given by $\lambda = - \tau_2 D_2/(\tau_1
D_1)$. Note that $\lambda$ is usually negative, but if it
is positive there no inconsistencies arising from a
negative $\tau_2$.

Inserting expression~(\ref{simpleQa}) into the time-wise
Fourier transformed equations of motion for $u_a$ and
$u_z$, we obtain
\begin{subequations}
\label{solidEOMs}
\begin{align}
\label{solidEOMsa} - \rho \omega^2 u_a &=   C_2 (\omega)
\partial_a u_{zz} + C_3 (\omega)\partial_a u_{ii} + 2 C_4
(\omega) \partial_b u_{ab}
\nonumber\\
& + C_5^R (\omega)  \partial_z u_{az} \, ,
\\
\label{solidEOMsz}
 - \rho \omega^2 u_z &=   [C_1 (\omega) + C_2(\omega)] \partial_z u_{zz}
+ [C_2 (\omega) + C_3 (\omega)] \partial_z u_{ii}
\nonumber\\
&+ C_5^R (\omega) \partial_b u_{bz}
\end{align}
\end{subequations}
with $C_1 (\omega)= C_1 - i\omega \nu_1$, $C_2 (\omega)=
C_2 - i\omega \nu_2$ etc.\  and
\begin{align}
\label{dynRen} C_5^R (\omega)&= C_5 - i \omega \left( \nu_5
+ \frac{\lambda^2}{2\, \Gamma} \right) - \frac{D_2^2}{2\,
D_1} \,  \, \frac{(1 - i \omega \tau_2)^2}{1-i \omega
\tau_1}
\nonumber \\
&= C_5^R - i\omega \nu_5^R + O \left( \omega^2 \right) \, ,
\end{align}
where
\begin{align}
\nu_5^R  =\nu_5  + \frac{\lambda^2}{2\, \Gamma} \, \left(
1- \frac{\tau_1}{\tau_2} \right)^2 .
\end{align}
Note that Eqs.~(\ref{solidEOMs}) are identical in form to
Eq.~(\ref{finalSimpleEOM}) for uniaxial solids with $K_1^R
= K_3^R = 0$.  Note also that we can identify $\eta_1 =
\nu_1$ and so on. Indicating the consistency of our two
approaches, we can identify the renormalized viscosities
$\eta_5^R$ and $\nu_5^R$. As pointed our earlier, Rouse
modes will be important for $\omega \tau_R \geq 1$ and $\omega \tau_\Gamma \gg 1$ leading
to a frequency dependence of the viscosities. In this
regime we have to let $\nu_i \to \nu_i (\omega)$ and $\Gamma \to \Gamma (\omega)$ . For $\omega
\tau_R \gg 1$ and $\omega \tau_\Gamma \gg 1$, in particular, the viscosities are
proportional to $\sqrt{\omega}$~\cite{rubinstein_colby_2003}.

Before we move on we point out that
Eq.~(\ref{effectiveEOMs}) become identical to the equations
of motion by TW if we take the incompressible limit and if
we neglect the Frank energy, provided of course, that we
take into account that somewhat different conventions are
used and provided that corresponding quantities are
identified properly. A detailed comparison to TW at the
level of final results for the modes will be given in
Sec.~\ref{compTW}.

\subsection{Mode structure}

\subsubsection{The $t$-direction}

By applying $\varepsilon_{ba} \, q_b$ (including the
summation over $a$) to both sides of the Fourier
transformed version of Eq.~(\ref{effectiveEOMsa}) we obtain
a diagonalized equation of motion for $Q_t$ that is readily
solved with the result
\begin{align}
\label{EOMQt} Q_t = -\frac{D_2}{D_1} \, \frac{1 - i
\omega\tau_2   + q^2 K_t/D_2}{1 - i \omega \tau_1 + q^2 K_t
/D_1} \, \frac{1}{2} \, i q_z u_t \, ,
\end{align}
where we have used the abbreviated notation
\begin{align}
q^2 K_t = K_2 q_\perp^2 + K_3 q_z^2 \, .
\end{align}
Application of the same procedure to
Eq.~(\ref{effectiveEOMsb}) yields
\begin{align}
\label{EOMut} - \omega^2 \rho u_t &= \frac{\lambda+1}{2 \,
\Gamma} \, i \omega q_z \left[  i Q_t + \frac{\lambda}{2}
\, q_z u_t\right]  - C_4 (\omega) \, q_\perp^2 u_t
\nonumber \\
& + \frac{D_2 - D_1}{2} \, i q_z Q_t - \frac{1}{2} \left[
C_5 (\omega) - \frac{D_2}{2}  \right] \, q_z^2 u_t \, .
\end{align}
$u_t$ is a purely transverse displacement that does not
couple to any displacement in the plane containing
$\brm{q}$. As a result, the elastic constants $C_2$ and
$C_3$, which couple to $u_l$, do not appear in this
equation of motion.

Next we insert Eq.~(\ref{EOMQt}) into (\ref{EOMut}) which
gives an effective equation of motion for $u_t$ alone. This
equation of motion has solutions with $u_t \neq 0$ for
frequencies satisfying the secular equation
\begin{align}
\label{sect} & \bigg[ \omega^2 \rho  - C_4  (\omega) \,
q_\perp^2 +  \frac{\lambda(\lambda+1) \, i \omega - \lambda
\tau_2^{-1}- 2 \Gamma C_5 (\omega) }{4 \, \Gamma}  \, q_z^2
\bigg]
\nonumber\\
& \times D_1 \,  [ 1- i \omega \tau_1 + q^2 K_t/D_1 ]
\nonumber\\
&+ \frac{\lambda + 1 -\tau_1^{-1} - \lambda \tau_2^{-1}}{4
\, \Gamma} \, q_z^2 \,  D_2 \, [1 - i \omega\tau_2 + q^2
K_t/D_2] = 0\, .
\end{align}
Evidently this secular equation is too complicated to find
useful closed solutions for $\omega$. However, since we are
interested in the long-wavelength behavior, we can
perturbatively determine solutions in the form of a power
series in the wavevector.

Let us first consider the simplified case of vanishing
$C$'s and $D$'s. In this case we obtain as expected the
well known slow and fast $t$-direction modes of nematic
liquid crystals
\begin{subequations}
\label{tModesRed}
\begin{align}
\label{wtmRed} \omega_{t,s} &=  - i \left[ K_2 \, q_\perp^2
+  K_3 q_z^2 \right] \bigg\{ \Gamma
 +  \frac{(1+\lambda)^2 \, q_z^2}{2\, (\nu_5 q_z^2 + 2 \nu_4 q_\perp^2)} \bigg\}  \, ,
\\
\label{wt+-Red}
 \omega_{t,f} &=  -i \, \frac{2 \nu_4\,  q_\perp^2 +  \nu_5 \, q_z^2}{2 \, \rho}  \, ,
\end{align}
\end{subequations}
as well as the previously announced spurious zero-frequency
mode. In writing Eqs.~(\ref{tModesRed}) we have considered
as usual the limit $K \rho \Gamma/\nu \ll 1$, where $K$
stands symbolically for all the Frank constants and $\nu$
stands ambiguously for all the viscosities. In comparing
Eqs.~(\ref{tModesRed}) and (\ref{TModesRed}) to the
original results on nematic liquid crystals as given in
Ref.~\cite{forster&Co_71} one should be aware of slight
differences in the notations. If we wish to use the
notation of Ref.~\cite{forster&Co_71} we have to replace in
Eqs.~(\ref{tModesRed}) $\Gamma \to \gamma^{-1}$, $\nu_1 \to
2 \nu_1$, $\nu_4 \to \nu_2$, $\nu_5 \to 2 \nu_3$, $q_\perp
\to q_1$ and  $q_z \to q_3$.

Now to the full secular equation~(\ref{sect}). Solving this
equation perturbatively leads also to three modes, namely
one massive mode and two propagating modes. The massive
mode has the frequency
\begin{align}
\label{wtm} \omega_{t,m}  &= - i \tau_1^{-1} - i \Gamma
\left[ K_2 \, q_\perp^2 +  K_3 q_z^2 \right] + i  \, \frac{
\nu_5^R - \nu_5 }{2\, \rho} \, q_z^2 \, .
\end{align}
Note that Eq.~(\ref{wtm}) does not depend on $C_5^R$.
Hence, this mode is shared by soft and semi-soft NEs. As
long as the sound velocities of the soft modes are finite,
their frequencies are
\begin{subequations}
\label{SofttModes}
\begin{align}
\label{wt+-}
 \omega_{t,\pm}  =&  \pm  \sqrt{\frac{2C_4 \, q_\perp^2 + C_5^R \, q_z^2}{2\rho}}  -i \, \frac{2 \nu_4\,  q_\perp^2 +  \nu_5^R \, q_z^2}{4\rho} \, ,
\end{align}
\end{subequations}
in full agreement with Eq.~(\ref{fullt}) when $C_5^R =0$
and the identifications $\nu_4 = \eta_4$ and $\nu_5^R =
\eta_5^R$ are made. For soft NEs the soft $t$-modes become
diffusive if $q_\perp =0$. The frequencies of these
diffusive modes are easily identified with those given in
Eqs.~(\ref{diffusiveModestT}) and
(\ref{diffusiveModestTslowFast}).

\subsubsection{The $T$-direction}
Except for the $t$-direction an analysis of the modes is
prohibitively complicated unless one resorts to the
incompressible limit.  In this limit $C_3 = \infty$ and
$u_l =0$, and as a result, $C_2$, $C_3$, $\nu_2$ and
$\nu_3$ do not appear in the equations of motion for $u_T$.

We need the equations of motion for the $\perp$ direction
as intermediate results for studying the $T$-direction.
Applying $q_a$ (including the summation over $a$) to the
Fourier transformed counterparts of
Eqs.~(\ref{effectiveEOMsa}) and (\ref{effectiveEOMsb}) we
obtain
\begin{subequations}
\label{perpEqs}
\begin{align}
\label{EOMQperp} Q_\perp &= -\frac{D_2}{D_1} \, \frac{1 - i
\omega\tau_2   + q^2 K_\perp/D_2}{1 - i \omega \tau_1 + q^2
K_\perp /D_1} \, \frac{1}{2} \, i q_\perp u_z
\nonumber\\
& -\frac{D_2}{D_1} \, \frac{1 - i \omega\tau_2   - q^2
K_\perp/D_2}{1 - i \omega \tau_1 + q^2 K_\perp /D_1} \,
\frac{1}{2} \, i q_z u_\perp \, ,
\\
\label{EOMuperp} - \omega^2 \rho u_\perp& =
\frac{\lambda+1}{2\, \Gamma} \, i \omega q_z \left[  i
Q_\perp + \frac{\lambda}{2} \, ( q_\perp u_z + q_z
u_\perp)\right]
\nonumber\\
&- 2 C_4 (\omega) \, q_\perp^2 u_\perp+ \frac{D_2 - D_1}{2}
\, i q_z Q_\perp
\nonumber\\
&-  \frac{1}{2} \left[ C_5 (\omega) - \frac{D_2}{2} \right]
q_z ( q_\perp u_z + q_z u_\perp) \, ,
\end{align}
\end{subequations}
where we have used the shorthand
\begin{align}
q^2 K_\perp = K_1 q_\perp^2 + K_3 q_z^2 \, .
\end{align}
In addition to Eqs.~(\ref{perpEqs}) we also need the
Fourier transformed version of Eq.~(\ref{effectiveEOMsc})
in the incompressible limit,
\begin{align}
\label{EOMuz} - \omega^2 \rho u_z& = \frac{\lambda-1}{2\,
\Gamma} \, i \omega q_\perp \left[  i Q_\perp +
\frac{\lambda}{2} \, ( q_\perp u_z + q_z u_\perp)\right]
\nonumber\\
&-  C_1(\omega) \, q_z^2 u_z+ \frac{D_1 + D_2}{2} \, i
q_\perp Q_\perp
\nonumber\\
&-  \frac{1}{2} \left[ C_5 (\omega) + \frac{D_2}{2} \right]
q_\perp ( q_\perp u_z + q_z u_\perp) \, .
\end{align}

Next we  set up an equation of motion for $u_T$ that
depends on $u_\perp$, $u_z$ and $Q_\perp$. Then we
eliminate $Q_\perp$ with help of Eq.~(\ref{EOMQperp}).
Finally, we exploit that $u_\perp = - q_z u_T/q$ and $u_z =
q_\perp u_T/q$ in the incompressible limit. These steps
provide us with an effective equation of motion for $u_T$
alone. In order to allow for solutions $u_T \neq 0$ the
frequencies have to satisfy a condition analogous to
Eq.~(\ref{sect}). We opt not to write down this secular
equation here because it is rather lengthy and because it
can be obtained in a straightforward manner from the
ingredients given above.

We proceed as above and first consider the simplified case
of vanishing $C$'s and $D$'s. As anticipated we obtain a
spurious zero-frequency mode as well as the slow and fast
$T$-direction modes of nematodynamics,
\begin{subequations}
\label{TModesRed}
\begin{align}
\label{wTmRed} \omega_{T,s} &=  - i   \left[ K_1 \,
q_\perp^2 +  K_3 q_z^2 \right]
\nonumber \\
&\times \bigg\{  \Gamma + \frac{[q^2 - \lambda(q_\perp^2 -
q_z^2)]^2}{2 \nu_5 (q_\perp^2 - q_z^2)^2 + 4 (\nu_1 + 2
\nu_4) q_\perp^2 q_z^2}   \bigg\}   \, ,
\\
\label{wT+-Red} \omega_{T,f} &=  -i \, \frac{2(\nu_1 + 2
\nu_4)\,  q_\perp^2 q_z^2+  \nu_5 \, \left(q_\perp^2 -
q_z^2\right)^2}{2\, \rho\, q^2}  \, ,
\end{align}
\end{subequations}
where $K \rho \Gamma/\nu \ll 1$ is implied.

By perturbatively solving the full secular equation we
extract the three $T$-direction modes for NEs. We find one
massive mode
\begin{align}
\label{wTm} \omega_{T,m}  &= - \tau_1^{-1} - i \Gamma
\left[ K_1 \, q_\perp^2 +  K_3  \right] q_z^2
\nonumber\\
&+ i \,  \frac{ \nu_5^R - \nu_5 }{2\, \rho} \, \frac{\left(
q_\perp^2 - q_z^2 \right)^2}{q^2}
\end{align}
and two soft modes. For non-vanishing sound velocities the
soft modes have frequencies
\begin{align}
\label{wT+-}
 \omega_{T,\pm}  &=  \pm  \sqrt{\frac{2(C_1 +2C_4) \, q_\perp^2  q_z^2+ C_5^R  \left( q_\perp^2 - q_z^2 \right)^2}{2\rho \, q^2}}
 \nonumber \\
 &-i \, \frac{2(\nu_1 + 2 \nu_4)\,  q_\perp^2 q_z^2 +  \nu_5^R  \left( q_\perp^2 - q_z^2 \right)^2}{4\rho\, q^2}   \, .
\end{align}
When $C_5^R =0$ this result reduces, provided that the
viscosities are properly identified, to Eq.~(\ref{fullT}).
In the case of ideal soft elasticity the sound velocities
vanish if $q_\perp=0$ or $q_z =0$. For $q_\perp=0$ the
frequencies of the then diffusive $T$-modes are identical
to those for the diffusive $t$-modes, see
Eqs.~(\ref{diffusiveModestT}) and
(\ref{diffusiveModestTslowFast}). For $q_z=0$ we retrieve
Eqs.~(\ref{noIdea1}) and (\ref{noIdea2}).

\subsection{Comparison to TW}
\label{compTW}
Now we compare our findings to those by TW. First, we will
demonstrate that our equations of
motion~(\ref{effectiveEOMs}) are identical to the equations
of motion by TW if we restrict ourselves to the
incompressible limit and if we neglect the Frank energy.
Second, we will compare our final results for the modes to
those found by TW. Of course, we must take into account
differences in conventions and notations in these
comparisons. For guidance in identifying corresponding
quantities, see Table~\ref{tab:notation}.
\begin{table}
\caption{Correspondence between quantities used by TW and
quantities used in our work.}
\label{tab:notation}
\begin{tabular}{|c|c||c|c|}
\hline
TW & this paper & TW & this paper\\
\hline
$2\, C_1$ & $C_1$ & $2\, A_4$ & $\nu_4$\\
$2\, C_2$ & $C_2$ & $4\, A_5$ & $\nu_5 + \lambda^2/(2\Gamma)$\\
$2\, C_3$ & $C_3$ & $\gamma_1$ & $1/\Gamma$\\
$2\, C_4$ & $C_4$ & $\gamma_2$ & $-\lambda/\Gamma$\\
$4\, C_5$ & $C_4$ & $\omega$ & $-\omega$\\
$2\, A_1$ & $\nu_1$ & $\theta_1$ & $Q_2$\\
$2\, A_2$ & $\nu_2$ & $\theta_2$ & $-Q_1$\\
$2\, A_3$ & $\nu_3$ & $C_5^R (\omega)$ & $\hat{C}_5^R (\omega)$\\
\hline
\end{tabular}
\end{table}

In order to compare our equations of
motion~(\ref{effectiveEOMs}) to the equations of motion by
TW we rewrite Eqs.~(\ref{effectiveEOMsb}) and
(\ref{effectiveEOMsc}) as
\begin{align}
- \rho \partial_t^2 u_i = \partial_j \sigma_{ij} \, ,
\end{align}
where the $\sigma_{ij}$ are the components of the stress
tensor $\tens{\sigma}$. The specifics of the $\sigma_{ij}$
are easily gathered from Eqs.~(\ref{effectiveEOMsb}) and
(\ref{effectiveEOMsc}),
\begin{subequations}
\label{stressElements}
\begin{align}
\sigma_{ab}&= 2 \left[ C_4 + \nu_4 \partial_t \right] \,
u_{ab}\, ,
\\
\sigma_{zz}&=  \left[ C_1 + \nu_1 \partial_t \right] \,
u_{zz} \, ,
\\
\sigma_{az}&=  \left[ C_5 + \nu_5 \partial_t -
\frac{D_2}{2}  + \frac{\lambda (\lambda +1)}{2\Gamma} \,
\partial_t \right] \, u_{az}
\nonumber \\
&+ \frac{1}{2} \, \left[ D_2 - D_1 - \frac{\lambda
+1}{\Gamma} \, \partial_t \right] \, Q_a \, ,
\\
\sigma_{za}&=  \left[ C_5 + \nu_5 \partial_t +
\frac{D_2}{2}  + \frac{\lambda (\lambda -1)}{2\Gamma} \,
\partial_t \right] \, u_{az}
\nonumber \\
&+ \frac{1}{2} \, \left[ D_2 + D_1 - \frac{\lambda
-1}{\Gamma} \, \partial_t \right] \, Q_a \, ,
\end{align}
\end{subequations}
where we have taken the incompressible limit.
Equations~(\ref{stressElements}) show clearly that
$\tens{\sigma}$ is not symmetric and that its antisymmetric
part is
\begin{align}
\label{difference} \sigma_{az} - \sigma_{za} &=
\frac{1}{\Gamma} \left\{ [\tau_1^{-1} +  \partial_t] \, Q_a
- \lambda [ \tau_2^{-1} +  \partial_t]\, u_{az}  \right\}
\nonumber \\
&=  M_{ab} (\brm{\nabla}) \, \tilde{\Omega}_b
\end{align}
where we have used Eq.~(\ref{effectiveEOMsa}) to obtain the
last equality. Equation~(\ref{difference}) reveals that the
stress tensor $\sigma_{ij}$ defined in Eqs.~(\ref{stressElements}) is symmetric only when the Frank
energy can be ignored.  Of course, when the Frank energy is
included, it is always possible, following the procedures
in Refs.~\cite{forster&Co_71} and \cite{Forster1983}, to
define a symmetric stress tensor $\sigma_{ij}^S$ yielding
the same equations of motion as $\sigma_{ij}$.

If we neglect the Frank energy, our stress tensor becomes
identical to the symmetric stress tensor used by TW
provided that we take into account
Table~\ref{tab:notation}. Moreover, as easily can be
checked, Eq.~(\ref{effectiveEOMsa}) becomes identical to
the balance of torques equation under these circumstances.
Therefore, our equations of motion~(\ref{effectiveEOMs})
are equivalent to the equations of motion by TW provided
that we take the incompressible limit and neglect the Frank
energy.

At this point it is interesting to compare the stability
conditions which are implicit in the dissipation function.
Our equations imply in the incompressible limit that $\nu_1
\geq 0$, $\nu_4 \geq 0$ and $\nu_5 \geq 0$. While the first
two conditions can be readily identified with the
conditions $A_1 \geq 0$, $A_4 \geq 0$ in the work of TW,
the situation is less obvious for the last condition. Note
that we can re-express $\nu_5 \geq 0$ in terms of the
quantities used by TW as
\begin{align}
\nu_5 = 4 A_5 - \frac{\gamma_2^2}{2 \gamma_1} \geq 0 \, .
\end{align}
which has an identical counterpart in the work of TW.
Stated in terms of the relaxation times, this condition
requires that
\begin{align}
\tau_1 \, \tau_R \geq \tau_2^2
\end{align}
for ideally soft NEs.

Next we come to the comparison of the results for the
modes. Our first observation is here that the massive modes
$\omega_{t,m}$ and $\omega_{T,m}$ in unaccounted for by TW.
Note that the soft modes in the $t$- and in the
$T$-direction are referred to in TW as qSH waves and qSV
waves, respectively.

Let us first compare the findings regarding the
$t$-direction. To foster this comparison we recast our
result~(\ref{wt+-}) as
\begin{align}
\label{wtmRecast} \omega^2 = \frac{1}{2\rho} \left[  2 C_4
(\omega) \, q_\perp^2+ C_5^R (\omega) \, q_z^2 \right] \, .
\end{align}
To the order we are working, i.e.\ to order $O (q^3)$, the
solutions to Eq.~(\ref{wtmRecast}) and our $\omega_{t,m}$
coincide. Note that Eq.~(\ref{wtmRecast}) is essentially
identical to the $t$-direction secular equation for
uniaxial solids. This can easily be checked by starting
with Eq.~(\ref{solidEOMsa}) and by then switching to the
$t$-direction.

To make further contact to WT we eliminate the viscosities
in Eq.~(\ref{wtmRecast}) in favor of the relaxation time
$\tau_R  \approx \nu_1/C_1 \approx \cdots \approx \nu_5/C_5
\approx [\nu_5 + \lambda^2/(2\Gamma)]/C_5$. The subscript
$R$ indicates that $\tau_R \approx 10^{-5} - 10^{-6}s$ is
of the order of the Rouse time of the polymers constituting
the rubbery matrix. We obtain
\begin{align}
\label{wtmRecast2} \omega^2 = \frac{1}{2\rho} \left[  2 C_4
\, q_\perp^2+ \hat{C}_5^R (\omega) \, q_z^2 \right]
(1-i\omega \tau_R)\, ,
\end{align}
where
\begin{align}
\label{wtmRecast3} &\hat{C}_5^R (\omega) = \frac{C_5^R
(\omega)}{1-i\omega \tau_R}
\nonumber \\
&= C_5 - \frac{D_2^2}{2D_1} \, \frac{(1 - i \omega
\tau_2)^2}{(1-i \omega \tau_1)\, (1-i\omega \tau_R)} \, .
\end{align}
Taking into account Table~\ref{tab:notation} we see that
our $\hat{C}_5^R (\omega)$ is identical with the
renormalized form of $C_5$ found by TW and that
Eq.~(\ref{wtmRecast3}) is  in full agreement with the
dispersion relation of TW for the qSH waves.

Now to the $T$-direction. To order $O(q^3)$ our results for
$\omega_{T,\pm}$ coincide with the solutions of
\begin{align}
\label{wTpmRecast} \omega^2 = \frac{1}{2\rho} \left[  [ C_1
(\omega) + 2 C_4 (\omega) ] \, \frac{q_\perp^2 q_z^2}{q^2}
+ C_5^R (\omega) \, \frac{\left(q_\perp^2 -
q_z^2\right)^2}{q^2} \right] .
\end{align}
Note that Eq.~(\ref{wTpmRecast}) is essentially identical
to the $T$-direction secular equation for conventional
uniaxial solids. TW considered the $T$-modes only for $q_z
=0$. For $q_z =0$ Eq.~(\ref{wTpmRecast}) reduces to
\begin{align}
\label{wTpmsimple} \omega^2 = \frac{1}{2\rho} \hat{C}_5^R
(\omega) \, q_\perp^2 \, (1 - i\omega \tau_R)  .
\end{align}
Using Table~\ref{tab:notation} find that
Eq.~(\ref{wTpmsimple}) is in full agreement with the result
of TW for the qSV waves.

Before summarizing our findings,  we now briefly return to
the property of NEs that makes them, as pointed out by TW,
candidates for acoustic polarizers, viz.\ the large
difference in attenuation between the $t$- the $T$-modes in
the symmetry direction where $q_z=0$. For the $t$-direction
it follows immediately from Eq.~(\ref{wtmRecast2}) that the
attenuation is proportional to $\omega$ if $C_5^R =0$ and
$q_z=0$. Equation~(\ref{wTpmsimple}) has two solutions for
$C_5^R =0$, namely $\omega =0$ and, with the proper
identification $\nu^R_5 = \eta^R_5$,  the fast diffusive
mode Eq.~(\ref{fastT}). If the Frank energy is taken into
account, the $\omega =0$ mode becomes the slow diffusive
mode of Eq.~(\ref{slowT}). Due to their diffusiveness, the
attenuation of these $T$-modes is proportional to
$\sqrt{\omega}$ and hence much larger for low frequencies
than the attenuation of the propagating $t$-mode. This
large difference in attenuation can be used, in principle,
to split the $T$-modes from the $t$-modes.

\section{Summary}
\label{concludingRemarks}

Nematic elastomers exhibit the remarkable phenomenon of
soft or semi-soft elasticity in which the effective shear
modulus $C_5^R$ for shears in planes containing the
anisotropy axis respectively vanishes or is very small.  In
this paper, we have explored the dynamical consequences of
this elasticity.  We derived dynamical equations, involving
only the displacement, valid in the hydrodynamic limit in
which frequencies and wavenumbers are respectively small
compared to all characteristic microscopic inverse times
and lengths in the system, and we determined that their
mode structure is identical to that of columnar liquid
crystals in the soft limit when $C_5^R = 0$.  We also used
the Poisson-bracket approach to derive dynamical equations,
which contain non-hydrodynamic modes, for both director and
displacement and verified that they reduced to those
derived by Terentjev, Warner, and coworkers
\cite{WarnerTer2003,terentjev&Co_NEhydrodyn} when
contributions from the Frank free energy can be ignored. We
analyzed the mode structure of these equations assuming a
single relaxation time for the director, which we took to
be longer than any other characteristic decay time such as
the Rouse time $\tau_R$. Our equations, however, permit the
introduction of frequency-dependent dissipative
coefficients valid at frequencies higher than these inverse
decay times.

Rheological experiments at zero wavenumber have reported
frequency-dependent storage and loss moduli that are in
agreement with the predictions of the semi-soft theory that
includes the director \cite{clarke&C0_2001}.  It would be
interesting to map out the modes of nematic elastomers
directly using light scattering.  It may, however, be
difficult to access the true hydrodynamic limit because it
applies in current NEs only for frequencies of order 100 Hz
or less and because inherent sample inhomogeneities may
lead to extra scattering that could mask the signals of the
characteristic modes of a homogeneous system.

\begin{acknowledgments}
We gratefully acknowledge support by the Emmy
Noether-Programm of the Deutsche Forschungsgemeinschaft
(OS) and the National Science Foundation under grant DMR
00-95631(TCL).
\end{acknowledgments}


\begin{thebibliography}{}
\bibitem{footnote1}Poisson brackets of coarse-grained variables require a
proper interpretation. Perhaps most naturally, they can be
regarded as coarse grained versions of the corresponding
microscopic Poisson brackets.
\end{thebibliography}
\end{document}